\documentclass[conference]{IEEEtran}
\IEEEoverridecommandlockouts

\usepackage[acronym]{glossaries}
\newacronym{acr:cv-qkd}{CV-QKD}{continuous-variable quantum key distribution}
\newacronym{acr:its}{ITS}{information-theoretic security}
\newacronym{acr:snr}{SNR}{signal-to-noise ratio}
\newacronym{acr:skr}{$\mathrm{SKR}$}{secret key rate}
\newacronym{acr:b2b}{B2B}{back-to-back}
\newacronym{acr:voa}{VOA}{variable optical attenuator}
\newacronym{acr:fer}{FER}{Frame error rate}
\newacronym{acr:ber}{BER}{bit error rate}
\newacronym{acr:ldpc}{LDPC}{low-density parity-check}
\newacronym{acr:bi-awgn}{BI-AWGN}{binary-input additive white Gaussian noise}
\newacronym{acr:met}{MET}{Multi-edge type}
\newacronym{acr:ir}{IR}{information reconciliation}
\newacronym{acr:dsp}{DSP}{digital signal processing}
\newacronym{acr:rl-ldpc}{RL-LDPC}{raptor-like low-density parity-check}
\newacronym{acr:dac}{DAC}{digital-to-analog converter}
\newacronym{acr:adc}{ADC}{analog-to-digital converter}
\newacronym{acr:pa}{PA}{Privacy amplification}
\newacronym{acr:qkd}{QKD}{Quantum key distribution}
\newacronym{acr:LLR}{LLR}{Log-likelihood ratio}
\newacronym[plural=CNs,firstplural=Check Nodes (CNs)]{acr:CN}{CN}{check node}
\newacronym{acr:VN}{VN}{variable node}
\newacronym{acr:MDR}{MDR}{multidimensional reconciliation}
\newacronym{acr:CRC}{CRC}{cyclic redundancy check}
\newacronym{acr:spa}{SPA}{sum-product algorithm}
\newacronym{acr:qrng}{QRNG}{quantum random number generator}

\usepackage{cite}
\usepackage{amsmath,amssymb,amsfonts}
\usepackage{algorithmic}
\usepackage{graphicx}
\usepackage{textcomp}
\usepackage{xcolor}
\usepackage{url}
\usepackage{wasysym}

\usepackage{amsmath,amssymb}
\usepackage{pgfplots}
\pgfplotsset{compat=1.18}
\usepgfplotslibrary{colormaps}
\usetikzlibrary{spy}
\usetikzlibrary{shapes.geometric}
\usepackage{subcaption}
\usepackage{graphicx}
\usepackage{caption}
\usepgfplotslibrary{groupplots}

\usepgfplotslibrary{external} 
\usepackage{comment}
\usepackage{siunitx}
\usepackage{wrapfig}
\usepackage{mathtools}
\usepackage[cal=cm]{mathalfa}

\definecolor{KIT_blue}{RGB}{70,100,170}
\definecolor{KIT_green}{RGB}{0,150,130}
\definecolor{KIT_red}{RGB}{162,34,35}
\definecolor{KIT_orange}{cmyk}{0,.45,1,0}

\DeclareMathOperator{\arctanh}{tanh^{--1}}
\DeclareMathOperator{\sign}{sgn}

\def\BibTeX{{\rm B\kern-.05em{\sc i\kern-.025em b}\kern-.08em
    T\kern-.1667em\lower.7ex\hbox{E}\kern-.125emX}}
\begin{document}

\title{An Open-Source Library for Information Reconciliation in Continuous-Variable QKD\\
\thanks{This work was funded by the German Federal Ministry of Education and Research (BMBF) under grant agreement 16KISQ056~(DE-QOR).}
}

\author{\IEEEauthorblockN{Erdem Eray Cil and Laurent Schmalen}
\IEEEauthorblockA{ {Communications Engineering Lab}, {Karlsruhe Institute of Technology}\\
Karlsruhe, Germany \\
\texttt{erdem.cil@kit.edu}}
}

\maketitle

\begin{abstract}
This paper presents an easy-to-use open-source software library for continuous-variable quantum key distribution (CV-QKD) systems. The library, written in C++, simplifies the crucial task of information reconciliation, ensuring that both communicating parties share the same secret key despite the noise.  It offers a comprehensive set of tools, including modules for multidimensional reconciliation, error correction, and data integrity checks. Designed with user-friendliness in mind, the library hides the complexity of error correction, making it accessible even to users without knowledge of error-correcting codes.
\end{abstract}


\begin{figure*}
    \centering
    \includegraphics{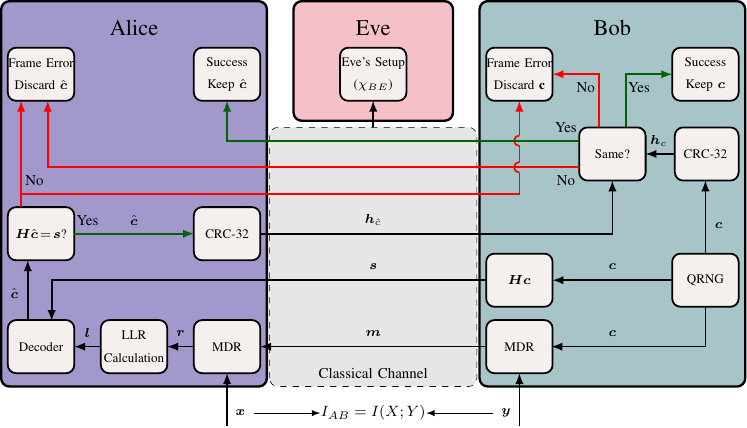}
    \caption{Illustration of the reverse reconciliation algorithm employing multidimensional reconciliation (MDR). QRNG, LLR, and $\boldsymbol{H}$  denote a random number generator, log-likelihood ratio, and parity check matrix of the error correcting code, respectively. Variables $\boldsymbol{x}$ and $\boldsymbol{y}$ represent the quantum states of Alice and Bob, respectively. }
    \label{fig:RR}
\end{figure*}

\section{Introduction}

The rapid progress in quantum computing threatens the security of many current encryption methods that rely on complex mathematical problems \cite{Djordjevic_book}.  \gls{acr:qkd} offers a powerful alternative, relying on the laws of physics to create and share secret keys for secure communication. This approach promises strong security, even against attacks from powerful quantum computers. 

Among the different \gls{acr:qkd} approaches, \gls{acr:cv-qkd} stands out due to its potential for long-distance communication and cost-effectiveness~\cite{208kmCV-QKD}. \gls{acr:cv-qkd}, in contrast to discrete-variable \gls{acr:qkd}, encodes information onto continuous properties of light, such as the amplitude and phase of electromagnetic field quadratures. This method often relies on readily available, off-the-shelf optical components, rendering it more practical and cost-effective.

A typical \gls{acr:cv-qkd} system comprises several stages to establish a secure key, which are briefly explained as follows:
\begin{enumerate}
    \item \textbf{State preparation and transmission:} One party, often referred to as Alice, prepares and transmits a series of quantum states. 
    \item \textbf{Reception and measurement:} The receiving party, typically called Bob, measures the received quantum states, often subject to noise introduced by the quantum channel and imperfections in the devices.
    \item \textbf{Digital signal processing:} Both Alice and Bob perform digital signal processing (DSP) on their respective data. DSP techniques are crucial for mitigating the impact of imperfections in the devices.
    \item \textbf{Post-processing:} The final stage involves \gls{acr:ir} to reconcile discrepancies between Alice's and Bob's data due to channel noise. \gls{acr:pa} extracts a final secret key from the reconciled data, accounting for potential information leakage during the reconciliation process. 
\end{enumerate}

Implementing a complete \gls{acr:cv-qkd} system from the ground up is a challenging endeavor that often requires expertise across multiple disciplines.  Recent initiatives, such as the open-source library QOSST \cite{QOSST}, aim to streamline this process by providing tools for various \gls{acr:cv-qkd} implementation stages. However, QOSST does not cover the crucial post-processing stage.

Recognizing this gap, we present an open-source C++ library\cite{lib} explicitly tailored for \gls{acr:ir} in \gls{acr:cv-qkd} systems. This library offers a rate-adaptive error-correcting code with rates ranging from $R=0.2$ down to $R=0.01$. This flexibility is especially beneficial for practical \gls{acr:cv-qkd} systems, where the signal quality can vary, requiring the use of codes with different rates. In addition to being rate-adaptive, the code also provides high reconciliation efficiencies, making the coding scheme included in the library suitable for middle- and long-range \gls{acr:cv-qkd} applications.

The library incorporates modules for \gls{acr:MDR}, decoding, and \gls{acr:CRC}, offering a comprehensive toolkit for \gls{acr:ir}. To make the library accessible to a broader audience, including researchers new to \gls{acr:cv-qkd}, we provide a user-friendly, black-box module. This module requires minimal user input (quantum states, noise variance, and desired code rate) and performs the entire \gls{acr:ir} process. Furthermore, for ease of use, the library includes Python bindings, allowing users to integrate it into their \gls{acr:cv-qkd} systems without needing in-depth knowledge of C++.

The paper is organized as follows: In Section~\ref{sec:IR}, we provide a concise overview of \gls{acr:ir} and the specifics of reverse reconciliation, including the \gls{acr:MDR}, error correction, and \gls{acr:CRC}. Section~\ref{sec:results} presents simulation results, highlighting the performance of our library.

\section{Information Reconciliation}
\label{sec:IR}
\gls{acr:ir} is a critical step in \gls{acr:cv-qkd} systems, enabling Alice and Bob to establish a shared secret key even in the presence of channel noise. This process involves exchanging information, often through a classical authenticated channel, to reconcile discrepancies between their correlated data. These discrepancies arise from the noisy nature of quantum channels and imperfections in the measurement devices. 

The choice of the \gls{acr:ir} method impacts the system's overall performance, particularly the achievable secret key rate and the maximum operational distance of the system. One key consideration is whether to employ direct error correction, where raw measurement data is reconciled, or to use an intermediate step like \gls{acr:MDR} to map the continuous data to a more manageable form. 

\gls{acr:MDR}, proposed in \cite{leverrierMultidimensionalReconciliationContinuousvariable2008}, enables us to transform the reconciliation problem into an equivalent problem over a \gls{acr:bi-awgn} channel, which is well-studied in classical coding theory. This transformation allows the use of efficient error correction codes designed for \gls{acr:bi-awgn} channels.

Another crucial aspect of \gls{acr:ir} is determining which party performs the reconciliation. In \textit{forward reconciliation}, Bob, the receiver, takes the lead in correcting errors based on the information received from Alice. Conversely, in \textit{reverse reconciliation}, Alice, the sender, assumes the role of error correction, using information fed back from Bob. Reverse reconciliation has been shown to be advantageous for achieving longer operational distances \cite{silberhornContinuousVariableQuantum}, and hence, implemented in the library.

\subsection{Reverse Reconciliation}

Figure~\ref{fig:RR} illustrates the reverse reconciliation process employed in our system. The following steps outline the process in detail:

\begin{enumerate}
    \item \textbf{Raw key generation:}  Using a \gls{acr:qrng}, Bob generates a sequence of random bits. This sequence serves as the basis for the raw key.
    \item \textbf{\gls{acr:MDR}:} Bob and Alice perform \gls{acr:MDR} on their respective quantum states, denoted by $\boldsymbol{y}$ and $\boldsymbol{x}$, respectively. \gls{acr:MDR} aims to map the continuous-variable quantum channel to a \gls{acr:bi-awgn} channel, simplifying the subsequent error correction process. 
    \item \textbf{\gls{acr:LLR} calculation and decoding:}  Based on the received quantum states and the outcome of \gls{acr:MDR}, Alice calculates the \gls{acr:LLR}s for each bit in the raw key. These \gls{acr:LLR}s are then fed into a channel decoder to correct potential errors introduced during transmission. %
    \item \textbf{Syndrome calculation and transmission:} Bob calculates the syndrome $\boldsymbol{s}$ of his generated raw key using the parity-check matrix $\boldsymbol{H}$ of the error-correcting code which is known to both parties. The syndrome is a compressed representation of the information required by the decoder to infer the correct raw key. Bob transmits this syndrome to Alice over the classical authenticated channel. 
    \item \textbf{Convergence check and \gls{acr:CRC} calculation:} Alice's decoder processes the calculated \gls{acr:LLR}s and the syndrome from Bob. If the decoder converges to a sequence that satisfies the parity check equations defined by the syndrome, Alice considers this sequence a candidate for the key material because it is a possible solution that satisfies the parity checks.  To verify that this potentially corrected sequence matches the one selected by Bob, Alice calculates the \gls{acr:CRC} of the sequence and sends it to Bob. 
    \item \textbf{\gls{acr:CRC} comparison and key storage/discarding:} Bob compares the received \gls{acr:CRC}  with the \gls{acr:CRC}  he computes for his version of the raw key. If both \gls{acr:CRC}s match, it indicates that both parties have successfully agreed on the same sequence, which is then stored for further processing in the privacy amplification stage. If the \gls{acr:CRC}s do not match, the reconciliation attempt for the current block of data is considered unsuccessful. The block is discarded, and the process can be repeated.  
\end{enumerate}

It is important to highlight that both the syndrome and the \gls{acr:CRC}  information transmitted from Bob to Alice represent information leakage. This leakage, while necessary for reconciliation, must be carefully accounted for during privacy amplification to ensure the security of the final secret key.

\subsection{Multidimensional Reconciliation}

\gls{acr:MDR} is a technique used to adapt the continuous-variable quantum channel for use with error correction codes primarily designed for \gls{acr:bi-awgn} channels. It involves rotating the quantum states in higher-dimensional space before transmission and reversing the rotation after reception. This process, as the dimensions of rotation ($d$) approach infinity ($d \to \infty$), effectively transforms the channel into a \gls{acr:bi-awgn} channel, making it compatible with conventional error-correction codes. 

However, practical implementations of \gls{acr:MDR} are limited to finite dimensions, typically $d=1$, $2$, $4$, and $8$. Consequently, the assumption of a perfect \gls{acr:bi-awgn} channel holds only approximately, and performance degrades as the dimension decreases. It is worth noting that, in principle, any orthogonal transformations could be used as long as they maintain information secrecy. However, the complexity of such general transformations often makes them impractical for \gls{acr:cv-qkd} systems~\cite{leverrierMultidimensionalReconciliationContinuousvariable2008}. Our library utilizes the Cayley-Dickson construction for implementing the rotations in \gls{acr:MDR} \cite{Milicevic_2018}.

\subsection{Error Correction}

Error correction is a crucial part of the \gls{acr:ir} process in \gls{acr:cv-qkd}, directly impacting the achievable secret key rate. To maximize the secret key rate, the error correction scheme must operate efficiently at low \gls{acr:snr}s typical for \gls{acr:cv-qkd}. This section elaborates on the specific error correction strategy used in our library, focusing on the code design and the decoding algorithm.

\subsubsection{Forward error correcting code}

\gls{acr:cv-qkd}  systems typically operate at much lower \gls{acr:snr}s than traditional communication systems. Consequently, the \gls{acr:ir} process requires error-correcting codes with significantly lower rates. \gls{acr:met}-\gls{acr:ldpc} codes are well-suited for such low to ultra-low code rate scenarios due to their excellent performance in these regimes.

Numerous \gls{acr:met}-\gls{acr:ldpc} code designs have been proposed for \gls{acr:cv-qkd}, each with its advantages and limitations~\mbox{\cite{Mani_thesis,9562244}}. However, to optimize performance over varying channel conditions, the code rate must be adaptable. While rate adaptation can be achieved using techniques like puncturing and shortening~\cite{Zhou2021}, these methods often result in performance degradation compared to codes explicitly designed for a specific rate~\cite{Cil24SPPCOM}. 

Our library addresses this challenge by incorporating a raptor-like \gls{acr:ldpc} code~\cite{Cil24SPPCOM}, optimized for a wide range of code rates, from $0.2$ down to $0.01$. This code, designed to perform well over a range of rates, eliminates the need for puncturing or shortening, thereby achieving higher efficiencies for the target rates. This code covers rates $R = 2 \cdot 10^4/(10^5 + i)$, where $i$ can take any integer value from $0$ to $1 \ 900 \ 000$.

One potential drawback of this specific code construction is that the block length is not independent of the rate. Choosing a particular rate implicitly determines the block length. This interdependence might appear restrictive, but it is not a major concern for \gls{acr:cv-qkd} systems. The high number of quantum states typically required for key distillation, driven by finite-size effects, implies that inherently long block lengths must be employed. Therefore, the fixed relationship between the rate and block length in our chosen code does not pose a practical limitation. 

Simulations demonstrate that this code can support distances up to \SI{110}{km} for rates between $0.2$ and $0.01$ \cite{ECOC}, making it particularly attractive for practical \gls{acr:cv-qkd} systems aiming for medium to long-distance communication.  This motivated us to include this code in our library.

\subsubsection{Decoding}

The \gls{acr:spa} has emerged as the dominant decoding algorithm for \gls{acr:ldpc} codes in \gls{acr:cv-qkd}. Its popularity stems from its near-optimal performance. The \gls{acr:spa} operates by iteratively exchanging messages, representing beliefs about the encoded bits, along the edges of a graph that represents the \gls{acr:ldpc} code.  

Our library's \gls{acr:spa} implementation utilizes \gls{acr:LLR}s as the message values to ensure numerical stability. The use of \gls{acr:LLR}s is particularly important for low code rates, where the magnitudes of probability values can become extremely small, leading to numerical instability if not handled appropriately.  

The library implements both flooding and layered scheduling for the \gls{acr:spa}. In flooding scheduling, all \glspl{acr:CN} update their outgoing messages in parallel based on incoming messages, followed by all \glspl{acr:VN} updating their outgoing messages. This process of \gls{acr:CN} updates followed by \gls{acr:VN} updates constitutes one iteration. In contrast, layered scheduling updates messages serially along specific connections in the graph representation of the code, rather than processing all nodes in parallel. This approach can lead to faster convergence by exploiting dependencies between messages and reducing redundant computations.

For completeness, we describe the flooding schedule for the \gls{acr:spa} with \gls{acr:LLR} messages. We use the following notation: \gls{acr:LLR}  values are represented by $L$. The synthetic channel \gls{acr:LLR} value ($\boldsymbol{l}$ in Fig.\ref{fig:RR}) for \gls{acr:VN} $i$ is denoted as $L_{\mathrm{ch},i}$.  In the context of an \gls{acr:ldpc} code with parity check matrix $\boldsymbol{H}$, the set \mbox{$\mathcal{N}(i)=\{ j: \boldsymbol{H}_{j,i} =1 \}$} represents the connection set of \gls{acr:VN}  $i$. Conversely, the set \mbox{$\mathcal{M}(j)=\{ i: \boldsymbol{H}_{j,i} =1 \}$} represents the connection set of \mbox{\gls{acr:CN} $j$}. Messages passed from \gls{acr:CN}/\gls{acr:VN} $j$ to \gls{acr:VN}/\gls{acr:CN}  $i$ are denoted by $L^{[\mathrm{c}]}_{i \gets j }$ and $L^{[\mathrm{v}]}_{j \to i }$, respectively. The block length of the code is represented by $N \in \mathbb{N}$, while $\boldsymbol{s}$ is the syndrome of the symbol sequence, with $s_j$ being its $j$th element.

\begin{enumerate}
    \item \textbf{Initialization:} For each \gls{acr:VN}  $i$, where $i \in \{ 1, \ldots, N \}$ and for each $j$ in $\mathcal{N}(i)$, initialize the message passed from \gls{acr:VN}  $i$ to \gls{acr:CN}  $j$ as $L^{[\mathrm{v}]}_{i \to j } = L_{\mathrm{ch},i}$.
    \item \textbf{\gls{acr:CN} update:} For each \gls{acr:CN}  $j$ and each \gls{acr:VN}  $i$ in $\mathcal{M}(j)$, the message passed from \gls{acr:CN}  $j$ to \gls{acr:VN}  $i$ is 
\begin{align}
   & L^{[\mathrm{c}]}_{i \gets j } = 2 s_j \arctanh \left( \hspace{-0.05cm}\prod_{\ell\in \mathcal{M}(j)\backslash \{i\}} \hspace{-0.4cm} \tanh\left(\frac{L^{[\mathrm{v}]}_{\ell \to j }}{2}\right) \right) . \label{eqn:full_SPA}
\end{align}
    \item \textbf{\gls{acr:VN} update:} For each \gls{acr:VN}  $i$ and each \gls{acr:CN}  $j$ in $\mathcal{N}(i)$, the message from \gls{acr:VN}  $i$ to \gls{acr:CN}  $j$ is updated as:
\begin{align}
   & L^{[\mathrm{v}]}_{i \to j } = L_{\mathrm{ch},i} +\sum_{k\in \mathcal{N}(i)\backslash \{j \}} L^{[\mathrm{c}]}_{i \gets k } . \nonumber 
\end{align}
    \item \textbf{A posteriori \gls{acr:LLR}  calculation:} For each \gls{acr:VN}  $i$, where $i \in \{ 1 \ldots N \} $, the a posteriori \gls{acr:LLR}  is
\begin{align}
    L^{[\mathrm{total}]}_{i } &= L_{\mathrm{ch},i} +\sum_{k\in \mathcal{N}(i)} L^{[\mathrm{c}]}_{i \gets k } . \nonumber
\end{align}
    The decoded bits are obtained by hard-deciding on these a posteriori \gls{acr:LLR}  values: 
    \begin{align}
        \hat{x}_i &= \frac{1- \sign{(L^{[\mathrm{total}]}_{i })}}{2}. \nonumber
    \end{align}
    \item \textbf{Stopping criteria:} The decoding process terminates if either of the following conditions is met:
    \begin{itemize}
        \item \textbf{Valid codeword:} The decoded sequence $\boldsymbol{\hat{x}}$, formed by concatenating the decoded bits as \mbox{$\boldsymbol{\hat{x}} = (\hat{x}_1, \ldots, \hat{x}_N )$}, satisfies the parity check equation: $\boldsymbol{H\hat{x}}^T=\boldsymbol{s}$.
        \item \textbf{Maximum iterations:} The number of decoding iterations reaches a predefined limit.
    \end{itemize}
     If neither condition is met, the algorithm proceeds back to Step 2, the \gls{acr:CN}  update.
\end{enumerate}

The calculation of the $\tanh$ and $\arctanh$ functions in \eqref{eqn:full_SPA} can be computationally intensive. To address this, our library provides an option to store the values of these functions in a lookup table, which can be pre-computed. During the decoding process, the decoder can retrieve the function values from the table instead of recalculating them, thereby speeding up the decoding process.

\subsection{Cyclic Redundancy Check}

A \gls{acr:CRC}, a widely used error-detection mechanism, is employed to ensure the integrity of data transmitted over a channel. It operates by generating a short checksum from a data sequence using polynomial division. By comparing the checksums calculated independently by the sender and receiver, any discrepancy indicates an error during transmission. 

In the context of our library, the \gls{acr:CRC} plays a crucial role in verifying if the decoder has converged to the correct codeword, that is, if the reconciled sequence at the receiver matches the original raw key generated by the sender. A \gls{acr:CRC}  mismatch indicates a decoding failure. Upon detecting a \gls{acr:CRC}  error, the current block of data is discarded. 

Our library implements a \gls{acr:CRC}-32 check for error detection. This choice is motivated by observations made in \cite{Milicevic_2018}, where a $32$-bit \gls{acr:CRC}  was found to be sufficient for detecting mis-decoded words in \gls{acr:cv-qkd} information reconciliation.

\begin{figure*}[hbtp] 
    \includegraphics{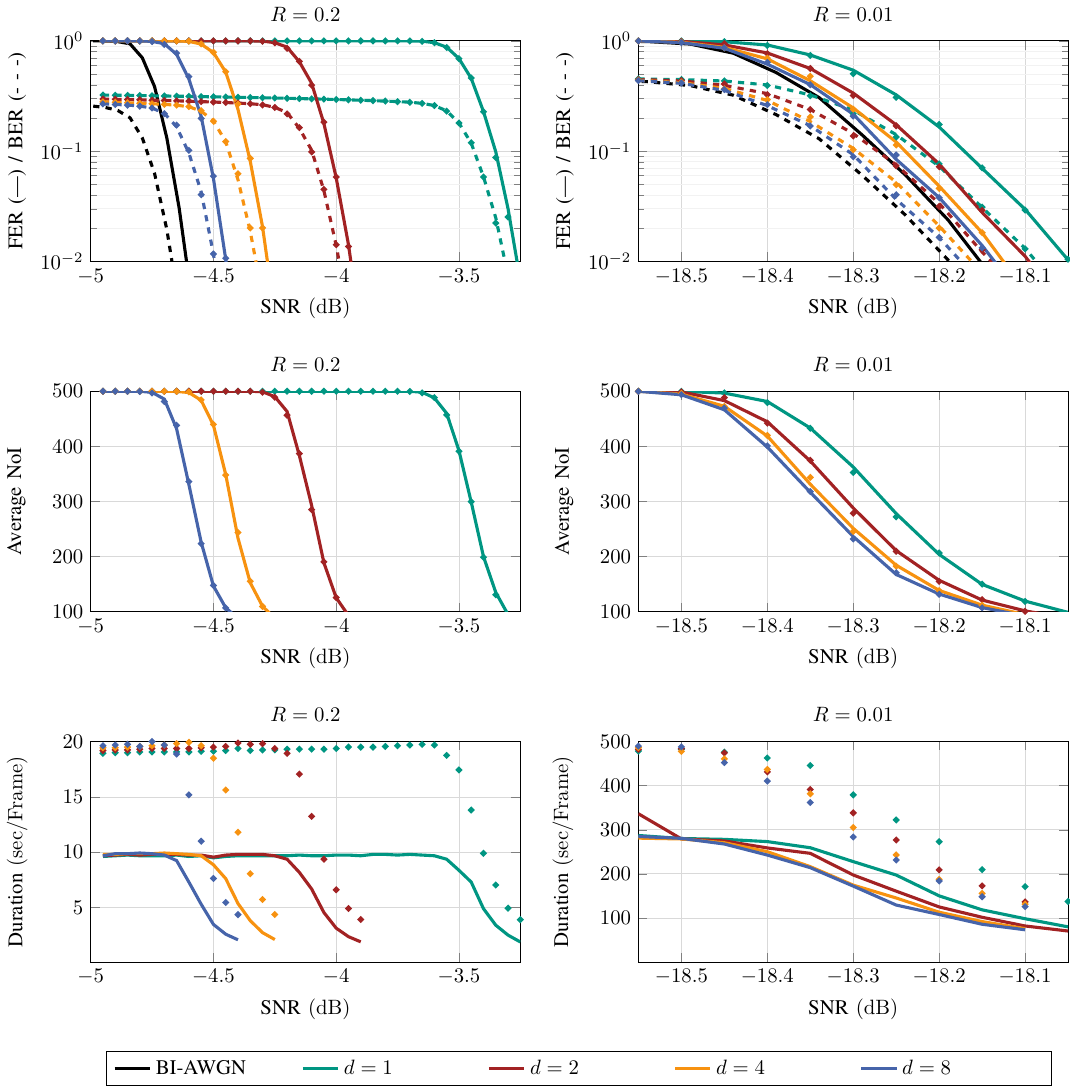}
 \caption{ Performance of the rate-adaptive LDPC code for code rates $R = 0.2$ (left) and $R = 0.01$ (right) and reconciliation dimensions ($d$) of 1, 2, 4, and 8. Frame error rate (FER), bit error rate (BER), average number of decoding iterations (NoI), and decoding duration per frame per CPU core are plotted as a function of the signal-to-noise ratio (SNR). Performance over a binary-input additive white Gaussian noise (BI-AWGN) channel is shown for comparison. Solid and dashed lines~ \mbox{(--- $/$ - - -)} represent decoders employing lookup tables, while markers ({\textbullet{}})  indicate direct calculation. Simulations were performed on an AMD EPYC\textsuperscript{TM} 7713P 64-Core Processor.}
 \label{fig:opt_results_3_figure}
\end{figure*}

\section{Results}
\label{sec:results}

To assess the performance of our open-source library, we conduct simulations emulating the reverse reconciliation scenario illustrated in Fig.~\ref{fig:RR}. We employ Gaussian-modulated coherent states and heterodyne detection. The \gls{acr:snr} is defined as  $1/\sigma_n^2$ of the quantum measurements. These measurements are subject to Gaussian noise. To accelerate the simulations, we design the use case to decode multiple frames concurrently, leveraging the processing power of multi-core CPUs. 

The simulations are conducted for different code rates and \gls{acr:MDR}  dimensions to assess the performance across various operating points relevant to \gls{acr:cv-qkd}.  We focus on two code rates, $R=0.01$ and $R=0.2$, representing ultra-low and low rate regimes, respectively. For each rate, simulations are performed for \gls{acr:MDR}  dimensions of  $d=1$, $2$, $4$, and $8$. 

Furthermore, we evaluate the impact of using a lookup table for the $\tanh$ and $\arctanh$ functions in the \gls{acr:spa} decoder.

For these simulations, we use \gls{acr:spa} decoding with a flooding schedule. The performance is evaluated using four metrics: \gls{acr:fer}, \gls{acr:ber}, average number of decoding iterations (NoI), and decoding duration of one decoding frame per CPU core.

Figure~\ref{fig:opt_results_3_figure} presents the simulation results, illustrating the interplay between code rate and \gls{acr:MDR}  dimension. The \gls{acr:fer}/\gls{acr:ber}  curves reveal a consistent trend: increasing the \gls{acr:MDR}  dimension, $d$, leads to improved decoding performance, particularly notable for the higher code rate, $R=0.2$. For instance, at $R=0.2$, using $d=8$ provides an \gls{acr:snr} gain of approximately \SI{1.5}{dB} compared to $d=1$ for an \gls{acr:fer}  target of $0.1$. However, this gain diminishes as the code rate decreases. For the ultra-low rate of $R=0.01$, the performance difference between $d=1$ and $d=8$ shrinks to about \SI{0.1}{dB} at the \gls{acr:fer} of $0.1$.

Examining the curves for $R=0.01$, it becomes evident that using $d=8$ pushes the performance close to the theoretical limit achievable over a true \gls{acr:bi-awgn} channel. This suggests that for ultra-low rates, increasing the \gls{acr:MDR}  dimension beyond $8$ offers diminishing returns. However, for $R=0.2$, a performance gap of roughly \SI{0.2}{dB} persists between $d=8$ and the theoretical \gls{acr:bi-awgn} limit at \gls{acr:fer}  $0.1$. This observation indicates that further performance improvements for higher code rates might be achievable by employing \gls{acr:MDR}  with dimensions greater than $8$.

Comparing the decoding durations of the real-time calculation versus the lookup table approach reveals a significant difference. Utilizing a lookup table for $\tanh$ and $\arctanh$ functions results in a remarkable $2\times$ speedup without compromising decoding performance. Both approaches yield identical decoding performances and require almost the same number of decoding iterations. This underscores the effectiveness of the lookup table method. 

\section{Conclusion}

This paper presents an open-source C++ library built for \gls{acr:ir}  in \gls{acr:cv-qkd} systems. One of the library's core strengths lies in its implementation of a rate-adaptive forward error correction scheme based on raptor-like \gls{acr:ldpc} codes. This allows the library to address a wide range of code rates, from $R=0.2$ down to $R=0.01$, with high reconciliation efficiency. This flexibility renders the library suitable for various \gls{acr:cv-qkd} setups, particularly those targeting middle to long-range distances.

The library offers a set of functionalities, including modules for \gls{acr:MDR}, high-performance \gls{acr:spa} decoding with both flooding and layered scheduling, and \gls{acr:CRC} for error detection. To facilitate broader adoption, the library includes a user-friendly, black-box module, simplifying the integration into experimental \gls{acr:cv-qkd} systems, even for users without specialized expertise in channel coding. Python bindings further enhance the library's accessibility, allowing for convenient integration into existing \gls{acr:cv-qkd} frameworks.

 \bibliographystyle{IEEEtran}
 \bibliography{IEEEabrv,biblio}

\end{document}